# A comparative study of HoSn$_{1.1}$Ge$_{0.9}$ and DySn$_{1.1}$Ge$_{0.9}$ compounds using magnetic, magneto-thermal and magneto-transport measurements

Sachin Gupta,[1] V. R. Reddy,[2] G. S. Okram[2] and K. G. Suresh[1*]

[1]Department of Physics, Indian Institute of Technology Bombay, Mumbai-400076, India

[2]UGC-DAE Consortium for Scientific Research, Khandwa Road, Indore-452017, India

## Abstract

Polycrystalline HoSn$_{1.1}$Ge$_{0.9}$ and DySn$_{1.1}$Ge$_{0.9}$ compounds have been studied by means of different experimental probes. Both the compounds are antiferromagnetic and show metamagnetic transition at low temperatures. HoSn$_{1.1}$Ge$_{0.9}$ shows a sign change in magnetocaloric effect (MCE) and magnetoresistance (MR) with field, which is attributed to the metamagnetic transition. DySn$_{1.1}$Ge$_{0.9}$ shows characteristics of a typical antiferromagnet, as evidenced by magnetization, MCE and MR data. $^{119}$Sn Mössbauer studies show hyperfine splitting at low temperatures, consistent with magnetization data. Thermoelectric power and resistivity measurements reveal metallic behavior in these compounds. Magnetic, magnetocaloric and the magnetoresistance data clearly show that the antiferromagnetic coupling in DySn$_{1.1}$Ge$_{0.9}$ is stronger than in HoSn$_{1.1}$Ge$_{0.9}$.

*Corresponding author email: *suresh@phy.iitb.ac.in*

# 1. Introduction

RSn$_{1+x}$Ge$_{1-x}$ (R= rare earth element) series is found to be interesting in respect of its magnetic and electrical properties. First time, Tobash et al. [1] synthesized the compounds in RSn$_{1+x}$Ge$_{1-x}$ (R=Y, Gd-Tm; x=0.15) series from the constituent elements using high-temperature reactions and molten Sn as a metal flux and studied structural, magnetic and electrical properties of the compounds of the RSnGe (now onwards RSn$_{1+x}$Ge$_{1-x}$ will be referred to as RSnGe throughout this paper) and found that all the compounds in the series are iso-structural and crystallize in the orthorhombic crystal structure. These compounds show antiferromagnetic ordering at low temperatures. Later on, Gill et al. [2] synthesized polycrystalline RSnGe (R=Tb-Er) samples using arc melting technique and performed neutron diffraction studies. This study has shown that magnetic structure of the compounds with R= Dy, Ho, and Er at low temperatures is described by the propagation vector (0.5, 0.5, 0) [2]. The magnetic structure in TbSnGe is sine wave modulated and is described by the propagation vector (0.4, 0, 0.5). Recently we have done detailed magnetic, magnetocaloric and magnetotransport studies on RSnGe (R=Gd, Tb, Er) compounds [3]. It has been found that among all the compounds, ErSnGe shows somewhat different behavior. It shows weak antiferromagnetism, which disappears at higher fields and can be attributed to a field induced metamagnetic transition. Occurrence of metamagnetic transition results in a large magnetic entropy change (near the magnetic ordering temperature), which is about 9.5 J/kg K for a field change of 50 kOe. This value is much larger than that of the other compounds of this series [3].

Motivated by the results obtained in some RSnGe (Gd, Tb, Er) compounds, we have studied some of these compounds in more detail with the view to understand other related properties with the help of some more characterization tools. In this paper, we discuss the magnetic, $^{119}$Sn Mössbauer, magnetocaloric, magnetotransport and thermoelectric properties of HoSnGe and DySnGe compounds.

## 2. Experimental details

The polycrystalline HoSnGe and DySnGe were synthesized by the arc melting the constituent elements. The purity of Sn and Ge was 99.99 % while for Ho and Dy it was 99.9%. As-cast samples were sealed in quartz tube in vacuum (10$^{-6}$ torr) and annealed for 7 days at 800 °C to remove any impurity phase (if any). After annealing, the phase purity of the samples was checked by room temperature x-ray powder diffraction pattern collected from X'PERT PRO diffractometer using CuKα radiation. The magnetization measurements, M(T) and M(H) were carried out using Quantum Design, Physical Property Measurement System (PPMS-6500). The electrical resitivity (ρ) measurements were carried out on a home-made system, by employing standard four probe technique applying an excitation current of 100 mA parallel to the magnetic field. Thermoelectric power (S) measurements were carried out using compacted pellets sandwiched between two Cu blocks, with reference to which the absolute $S$ was measured [4]. Transmission $^{119}$Sn Mössbauer measurements were carried out using a conventional constant-acceleration spectrometer equipped with WissEl velocity drive. The velocity scale was calibrated with a $^{57}$Co(Rh) source and a metallic iron foil, at room temperature. The $^{119}$Sn source was kept at room temperature while the temperature of the absorber, placed inside Janis helium cryostat, was varied. The Mössbauer spectrum was recorded at different temperatures so as to cover the transition temperature.

## 3. Experimental Results

The Rietveld analysis of room temperature x-ray powder diffraction (XRPD) shows that both the compounds crystallize in orthorhombic crystal structure with the centrosymmetric space group Cmcm (SG# 63). The lattice parameters obtained from the refinements are a=4.24(1) Å, b=16.14 (7) Å and c= 4.05 (1) Å for HoSnGe and a=4.25(6) Å, b=16.23 (3) Å and c= 4.05 (7) Å for DySnGe and are very close to reported values [1]. The XRPD pattern for HoSnGe as a representative plot is shown in Fig. 1.

Fig. 2 shows the temperature dependence of dc magnetic susceptibility on the left hand axis and inverse magnetic susceptibility along with Curie-Weiss fit on right hand axis for the title compounds. The susceptibility data show a cusp at low temperatures, indicating the antiferromagnetic ordering in these compounds. The fit yields the effective magnetic moment ($\mu_{eff}$) and paramagnetic Curie temperature ($\theta_p$) of 10.8 $\mu_B$/Ho$^{3+}$ and -10.5 K for HoSnGe and 11.1 $\mu_B$/Dy$^{3+}$ and -16.5 K for DySnGe, respectively. It may be noted that free ion magnetic moment for Ho$^{3+}$ and Dy$^{3+}$ are 10.6 $\mu_B$/Ho$^{3+}$ and 10.63 $\mu_B$/Dy$^{3+}$, respectively. Therefore, in the case of HoSnGe, $\mu_{eff}$ is close to the expected value, while DySnGe shows slightly higher than the expected value. The negative sign of $\theta_p$ confirms the antiferromagnetic ordering in both the compounds. The magnetic transition temperatures (Néel temperature, $T_N$) have been estimated from the temperature derivative of susceptibility ($d\chi/dT$) and are found to be 11 K and 16 K, which are close to the values reported in Ref. [1].

The field dependence of magnetization at different temperatures, with a temperature interval of 2 K and for fields up to 50 kOe is shown in Fig 3. It can be observed that at low temperatures (below $T_N$) both the compounds show field induced metamagnetic transition. One can note from Fig. 3 that above the metamagnetic transition, HoSnGe shows a curvature at higher fields, while the magnetization remains linear in the case of DySnGe. In HoSnGe, the magnetization at low temperatures (4 and 6 K) makes a crossover with increase in field. At low fields the magnetization increases with temperature whereas at higher field the trend is opposite, which manifests that in HoSnGe the metamagnetic transition changes its state from antiferromagnetic to a weakly ferromagnetic. DySnGe also shows field induced metamagnetic transition, but it appears that the magnetic state after the field induced transition is the same as before the transition (AFM→ AFM). One can see from Fig. 3 (b) that there is no crossover of magnetization with field in this case. Thus DySnGe shows antiferromagnetic ordering throughout the field range under study. The value of critical field ($H_C$) required for the metamagnetic transition has been estimated from the derivative of magnetization with respect to field ($dM/dH$) and found to be 26 kOe at 4 K and 38 kOe at 2 K for HoSnGe and DySnGe, respectively. The magnetization derivative plots for the title compounds are shown as insets in Fig. 3. The compounds do not show hysteresis and the saturation for fields up to 50 kOe.

The heat capacity data with and without field are shown in Fig. 4 for both the compounds. Both compounds show λ- shaped peak near the onset of the magnetic ordering in zero field heat capacity, which suggests the second order nature of the magnetic order-disorder transition. The application of field suppresses the peak as it broadens on the application of field in case of HoSnGe, suggesting weak metamagnetic transition (as revealed by the non saturation trend in the magnetization isotherms). Since the metamagnetic transition does not change the magnetic state in DySnGe after metamagnetic transition, the compound is antiferromagnetic, which is confirmed from the heat capacity data.

Magnetocaloric effect (MCE) in terms of the change in isothermal magnetic entropy has been estimated from magnetization data using Maxwell's relation, $\Delta S_M = \int_0^H \left( \frac{\partial M}{\partial T} \right)_H dH$, where M is the magnetization and H is the applied field. The temperature dependence of $-\Delta S_M$ is shown in Fig. 5. It is well known that ferromagnetic materials show positive MCE, while antiferromagnetic materials show negative MCE around their magnetic transition temperatures. One can see in Fig. 5 that both HoSnGe and DySnGe show positive magnetic entropy change (negative MCE) at low temperatures. But in HoSnGe, the sign of $\Delta S_M$ changes and shows a caret-like peak around its ordering temperature. This confirms the AFM-FM transition as seen from the magnetization data. As a reflection of the difference in the magnetization data, MCE behavior in DySnGe is different from that of HoSnGe. MCE in DySnGe remains predominantly negative throughout the ordered temperature regime. The small positive MCE is seen in the paramagnetic regime at high fields, which is expected for paramagnetic materials. A careful observation of the entropy change at low temperatures in HoSnGe reveals that the magnitude of the entropy change decreases with increase in field, which is the consequence of the aligning of the moments towards the field direction. From Fig. 5, it is clear that HoSnGe shows the conventional MCE with a maximum value of 6.3 J/kg K and DySnGe shows inverse MCE with a value of 5 J/kg K for the field change of 50 kOe around their ordering temperatures.

To study these systems further and to correlate magnetic, magnetocaloric and electrical properties, we have carried out resistivity measurements with and without fields. Zero field resistivity data are plotted in Fig. 6 for both the compounds. The resistivity shows positive temperature coefficient in the paramagnetic regime, reflecting the metallic nature of the compounds. There is a sharp change in resistivity below ordering temperature, which is expected in the ordered regime due to the loss of spin disorder contribution. The insets in Fig. 6 show the low temperature resistivity data in different fields. From the inset in Fig. 6(a), in HoSnGe, one can note that there is very little effect of magnetic field on the resistivity in the paramagnetic regime, but it is influenced by field near the ordering temperature. It is also worth noting from the insets that the nature of resistivity in DySnGe in presence of field is different from that in HoSnGe. In HoSnGe, on the application of field the resistivity decreases with field near $T_N$, while below $T_N$, the resistivity shows an increase with increase in field. In case of DySnGe, the resistivity shows an increase with the application of field near $T_N$ and the behavior remains same down to the lowest measured temperature. The differences in the resistivity behavior on the application of field in HoSnGe and DySnGe near $T_N$ arise due to differences in the field induced magnetic states.

The magnetoresistance (MR) for these compounds has been estimated from the field dependence of resistivity using the relation, MR=[$\rho$(H,T)- $\rho$(0,T)]/ $\rho$(0,T) (where H is the applied field) and is shown in Fig. 7. HoSnGe shows negative MR near the ordering temperature (10-20 K), which is attributed to the suppression of the disorder by the applied field. The magnitiude of the negative MR is negligible for the fields up to about 26 kOe, but shows a noticeable increase beyond this field. On the other hand, at low temperatures (4 and 8 K), it shows positive MR. It is interesting to note that the positive MR increases for fields up to about 26 kOe and then decreases with further increase in field. Therefore, the nature of MR in both the low and the high temperature regimes shows a change at about 26 kOe, which is close to the critical field for metamagnetic transition (calcualted from the magnetization isotherms). On the basis of the MR data of HoSnGe, it is clear that near ordering temperature, the field is able to break the antiferromagnetic coupling, which results in negative MR. However at lower temperatures (4 and 8 K), initially the applied field is not strong enough to break the anitiferrromagnetic coupling among the moments and hence the compound shows positive MR, which increases with field and changes sign at fields above $H_C$. A similar MR behavior was also seen in HoRhGe which is an antiferrromagnet, undergoing the metamagnetic transition [5]. On the other hand, DySnGe shows positive MR throughout the temperature range under study. Below the ordering temperature, the MR increases with field and shows saturation tendency near $T_N$, while it shows non-saturating behavior at low temperatures. The positive MR is usually expected for antiferromagnetic materials, in which the application of field causes some randomness of moments. Therefore, MCE and MR data clearly suggest that the antiferromagnetic ordering in DySnGe is stronger than that in HoSnGe. It can be noted from Fig. 7(a) that HoSnGe shows significant value of MR at 20 K (i.e., in paramagnetic regime). The negative MR in paramagnetic regime generally arises due to suppression of spin fluctuations on the application of field and shows $H^2$ dependence [6]. The quadratic field dependence of MR at 20 K confirms the suppression of spin fluctuations by the applied field. The hump in MR for both the compounds at 4 K may arise due to Indium (In) contacts, since Indium has its superconducting transition near 4 K.

The $^{119}$Sn Mössbauer spectra for both HoSnGe and DySnGe compounds in the paramagnetic as well as in the ordered regime are shown in Fig. 8. With the help of $^{119}$Sn Mössbauer spectroscopy, we try to understand the magnetism and temperature dependence of hyperfine parameters. One can see from Fig. 8 that, in paramagnetic regime (i.e. at 30 and 300 K), the observed spectra for both HoSnGe and DySnGe samples exhibit a quadrupole doublet as expected for non cubic point symmetry of the Sn site [7]. High temperature (T>T$_N$) $^{119}$Sn Mössbauer data is fitted with a doublet and the observed hyperfine parameters at room temperature viz., isomer shift (IS), $\delta_{IS}$= 2.14 ± 0.01 mm/s, width of the lines, (FWHM) = 0.87 ± 0.03 mm/s and quadrupole splitting (QS) of 1.14 ± 0.02 mm/s are the same (within experimental error) for both the compounds. In the ordered regime (i.e., at 5 and 10 K) the main doublet begins to split due to the magnetic ordering. This is similar to the result of Gurgul et al., for HoRhSn [8]. The width of the lines obtained from the high temperature data (30 K) was used for fitting the low temperature (T<T$_N$) magnetically split spectra. It may be noted that we have adopted the fitting formalism of Gurgul et al. [8], i.e., both polar angles θ and ϕ are kept as 90º. This is equivalent to assuming that the direction of the hyperfine field is parallel to the $c$-axis, which is also consistent with the neutron diffraction data [2], in which it has been reported that the moments are aligned along $c$-axis for both the these compounds. The observed hyperfine field values are shown in Tables 1 and 2. The hyperfine field arises due to the magnetic moment of the rare earth ion in both these compounds. From the tables one can note that the value of B$_{hf}$ in the case of DySnGe is higher than that of HoSnGe. The difference is mainly attributed to the difference in the magnetic ordering temperatures of Dy and Ho compounds.

**Table I. Hyperfine parameters obtained from the $^{119}$Sn Mössbauer spectra of HoSnGe compounds at different temperatures. $\chi^2 \cong 1$ indicates the reliability of the fit.**

| Temperature (K) | FWHM (mm/s) | IS (mm/s) | QS (mm/s) | B$_{hf}$ (Tesla) | $\chi^2$ |
|---|---|---|---|---|---|
| 300 | 0.87±0.03 | 2.14±0.01 | 1.14±0.02 | --- | 0.8 |
| 30 | 0.99±0.02 | 2.42±0.01 | 1.22±0.01 | -- | 1.23 |
| 10 | 0.99 | 2.46±0.01 | 1.26±0.01 | 0.73±0.02 | 1.08 |
| 5 | 0.99 | 2.43±0.01 | 1.32±0.02 | 1.02±0.02 | 1.10 |

**Table II. Hyperfine parameters obtained from the [119]Sn Mössbauer spectra of DySnGe compounds at different temperatures. $\chi^2 \cong 1$ indicates the reliability of the fit.**

| Temperature (K) | FWHM (mm/s) | IS (mm/s) | QS (mm/s) | B$_{hf}$ (Tesla) | $\chi^2$ |
|---|---|---|---|---|---|
| 300 | 0.88±0.02 | 2.14±0.01 | 1.14±0.01 | --- | 0.7 |
| 30 | 1.16±0.02 | 2.41±0.01 | 1.26±0.01 | -- | 0.93 |
| 10 | 1.16 | 2.39±0.02 | 1.78±0.03 | 1.24±0.03 | 0.99 |
| 5 | 1.16 | 2.41±0.01 | 1.70±0.02 | 1.23±0.01 | 1.4 |

Fig. 9 shows the temperature dependence of the Seebeck coefficient (S(T)) for HoSnGe and DySnGe compounds. The compounds show positive thermoelectric power in the whole temperature range of investigation, which indicates hole conduction in both of them [9]. The shape of thermoelectric power curve is more or less similar to that of some intermetallic compounds such as La$_7$Ni$_3$ [10] and Sm$_2$PdGe$_6$ [11]. The small values of S(T) are indicative of typical of metals, in agreement with that observed in resistivity. In HoSnGe, S(T) decreases with temperature exhibiting nearly linear behavior down to about 120 K. Below 120 K, the slope changes such that S reaches a minimum at about 55 K from which it increases quite abruptly until it attains a maximum near 11 K. In a material, as temperature is lowered, phonon flow also contributes dominantly to the flow of charge carriers, creating the so-called phonon drag peak or dip in the Seebeck coefficient [12,13]. This is assigned to the dip near 55 K. As is clear, very similar transport mechanism prevails in DySnGe sample also.

In the present samples, it is also interesting to note that while the resistivity for HoSnGe is slightly higher than that of DySnGe, thermoelectric power plots show the opposite behavior at low and high temperatures. Such scenario is considered to be beneficial for enhancing the thermoelectric figure of merit of the material [14]. To better understand the electrical transport in these materials, inset of Fig. 9 shows the ST vs. T$^2$ plots. We have fitted the high temperature region of these plots with the expression ST=aT$^2$+b, which takes into account the combined diffusion and phonon drag contributions (where *a* and *b* are charge carrier diffusion coefficient and phonon drag coefficient, respectively). The values of the coefficients obtained from the fittings are a=0.007 µV and b= -49.43 µV/K$^2$ for DySnGe and a= 0.004 µV and b=10.44 µV/K$^2$ for HoSnGe. The origin of different signs of b is not understood at present. This means that the charge carrier diffusion in DySnGe is nearly double in the former that of the later and the phonon drag is nearly five times that of HoSnGe, indicating a significant difference in the response of their electron and phonon transport mechanisms. Such a difference may be related to the stronger antiferromagnetic coupling strength in DySn$_{1.1}$Ge$_{0.9}$ than that in HoSn$_{1.1}$Ge$_{0.9}$.

## Conclusions

Polycrystalline HoSnGe and DySnGe crystallize in the orthorhombic crystal structure. Both the compounds have antiferromagnetic ground state and show metamagnetic transition. Application of field transforms the magnetic state from antiferromagnetic to weakly ferromagnetic in HoSnGe, as confirmed by magnetization, magnetocaloric and MR data. HoSnGe shows a sign change of both MCE and MR with change in temperature, while DySnGe shows negative MCE and positive MR, thereby reflecting the stronger antiferromagnetic coupling. Thermoelectric power and electrical resistivity measurements reveal metallic behavior in these compounds. $^{119}$Sn Mössbauer studies show hyperfine splitting below T$_N$ in both these compounds and a magnetic structure in which the R moments are parallel to the *c*-axis.

## Acknowledgments

SG would like to thank CSIR New Delhi for providing senior research fellowship. Authors thank Dr. R. Rawat, UGC-CSR Indore for using resistivity, Mössbauer and thermoelectric power facilities.

## References

[1] P. H. Tobash, J. J. Meyers, G. DiFilippo, S. Bobev, F. Ronning, J. D. Thompson, J. L. Sarrao, Chem. Mater. 20 (2008) 2151.

[2] A. Gil, B. Penc, S. Baran, A. Hoser, A. Szytuła, J. Solid State Chem. 184 (2011) 1631.

[3] S. Gupta, K. G. Suresh, A. K. Nigam, J. Magn. Magn. Mater. 342 (2013) 61.

[4] A. Soni, G. S. Okram, Rev. Sci. Instrum. 79, 125103 (2008).

[5] S. Gupta, K. G. Suresh, A. K. Nigam, Yu. V. Knyazev, Yu. I. Kuz'min, and A. V. Lukoyanov, J. Phys. D: Appl. Phys. 47 (2014) 365002.

[6] S. Gupta, R. Rawat and K. G. Suresh, Appl. Phys. Lett. 105 (2014) 012403.

[7] K. Łatka, R. Kmieć, J. Gurgul, A. W. Pacyna, M. Rams, T. Schmidt, R. Pöttgen, J. Magn. Magn. Mater. 301 (2006) 359.

[8] J. Gurgul, K. Łatka, A.W. Pacyna, C. P. Sebastian, R. Pöttgen, Intermetallics 18 (2010) 129.

[9] V. Goruganti, K. D. D. Rathnayaka, J. H. Ross, Jr., Y. Öner, C. S. Lue, Y. K. Kuo, J. Appl. Phys. 103 (2008) 073919.

[10] H. Kadomatsu, Y. Itoh, H. Fukuda, T. Tsutaoka, T. Tokunaga, J. Magn. Magn. Mater. 177 (1998) 1117.

[11] R. Troc, R. Wawryk, K. Gofryk, A. V. Gribanov and Yu. D. Seropegin, J. Phys.: Condens. Matter 23 (2011) 146001.

[12] F. J. Blatt, P. A. Schroeder, C. L. Foiles, D. Greig, Thermoelectric Power of Metals, Plenum Press, New York (1976).

[13] D. K. C. MacDonald, *Thermoelectricity: An Introduction to the Principles* John Wiley and Sons, New York (1962).

[14] G. J. Snyder and E. S. Toberer, Nature Mater. 7 (2008) 105.

## Figure captions

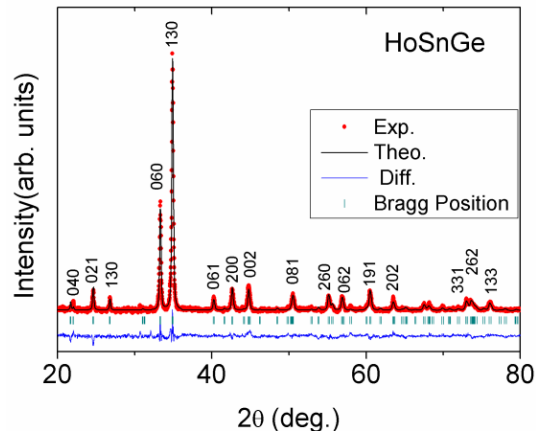

Fig. 1. XRPD pattern along with Rietveld analysis for HoSnGe as a representative plot. The blue line in the pattern shows the difference between the experimental and the theoretical patterns.

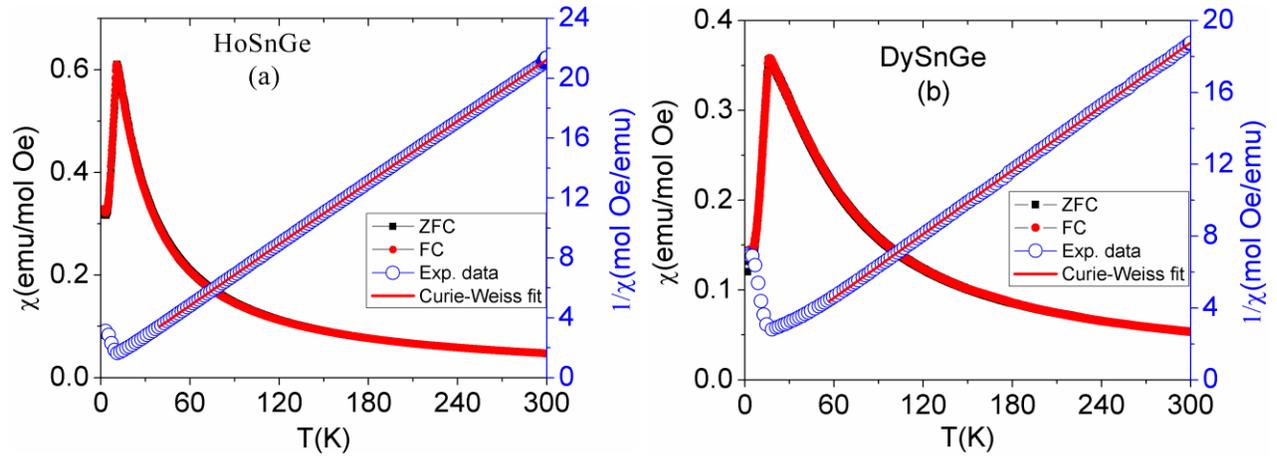

Fig. 2. Temperature dependence of magnetic susceptibility in 500 Oe (left-hand panel) for HoSnGe and DySnGe. The right hand panel shows the Curie-Weiss fit to the inverse susceptibility data.

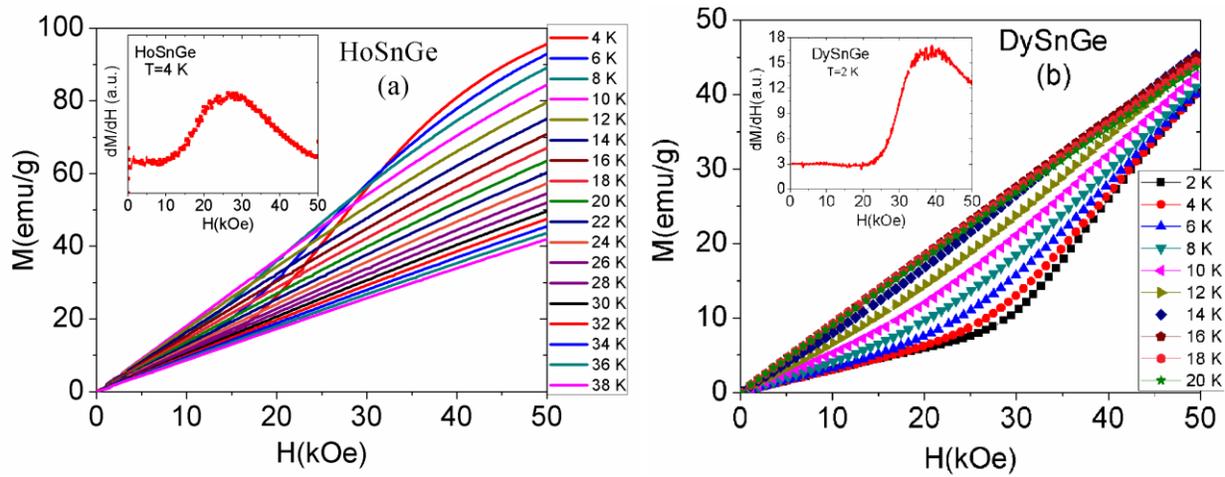

Fig. 3. Field dependence of magnetization in HoSnGe and DySnGe at different temperatures. The inset shows field deriveative of magnetization.

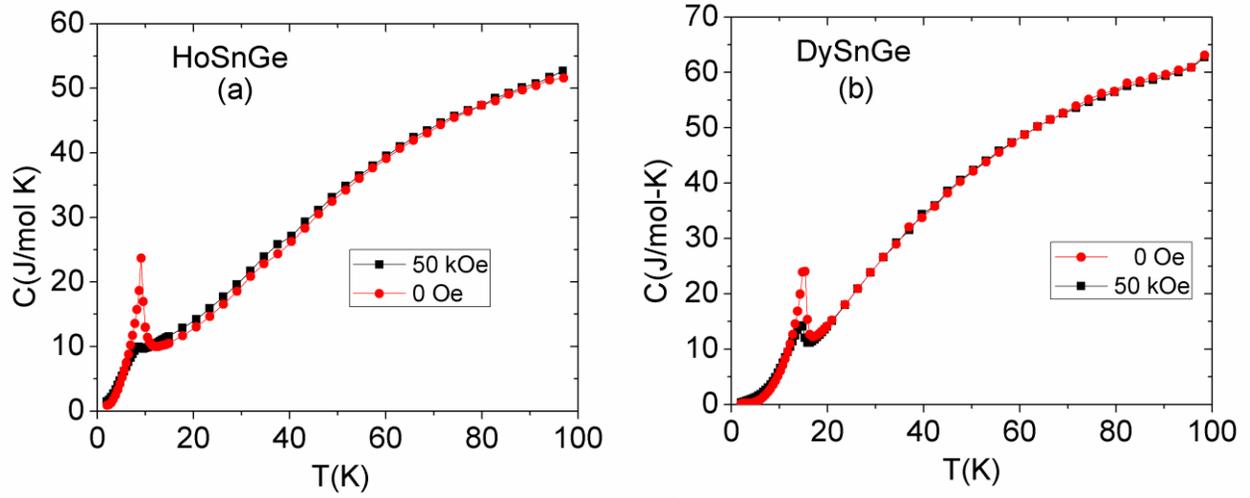

Fig. 4. Temperature dependence of the heat capacity in HoSnGe and DySnGe in zero and 50 kOe fields.

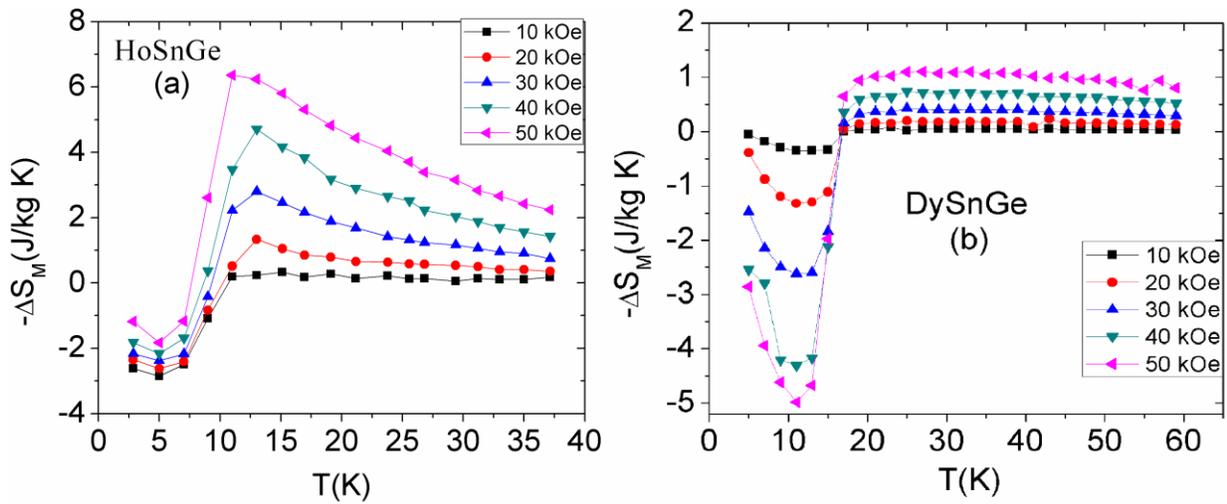

Fig. 5. Temperature dependence of isothermal magnetic entropy change for different fields in HoSnGe and DySnGe.

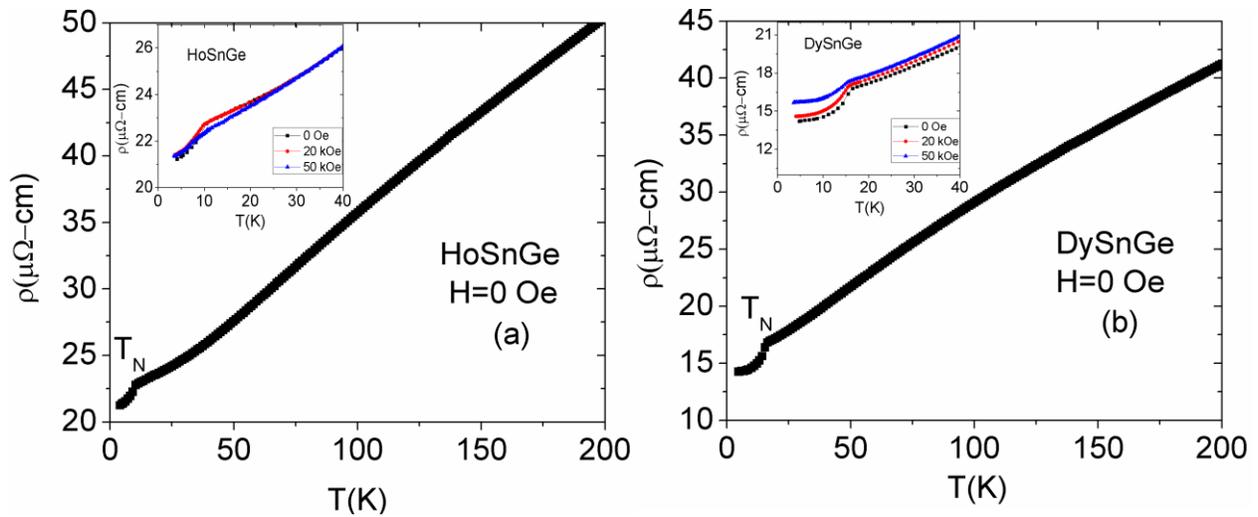

Fig. 6. Temperature dependence of electrical resistivity in HoSnGe and DySnGe. Inset shows the low temperature electrical resistivity data in 0, 20 and 50 kOe fields.

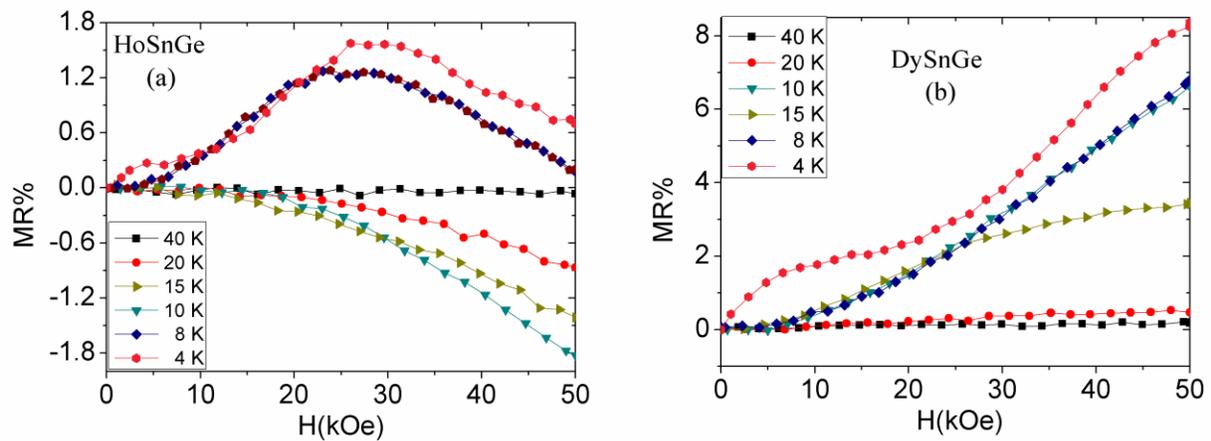

Fig. 7. Field dependence of magnetoresistance in HoSnGe and DySnGe at selected temperatures.

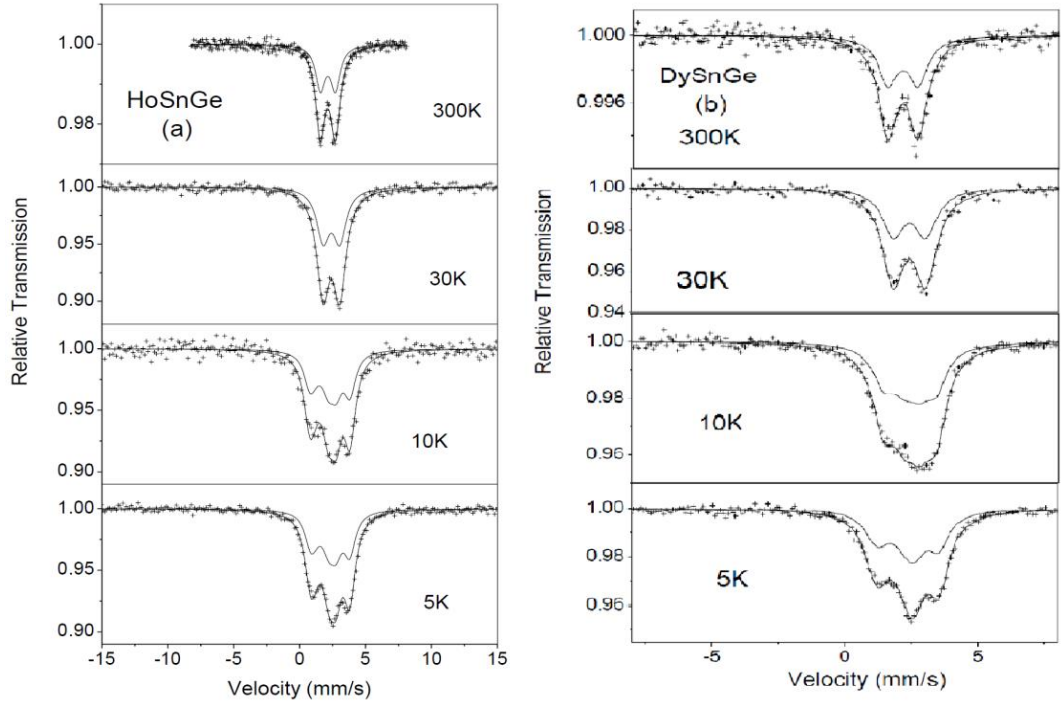

Fig. 8. $^{119}$Sn Mössbauer spectra obtained at different temperatures for (a) HoSnGe and (b) DySnGe. The solid line is the least square fit to the experimental data.

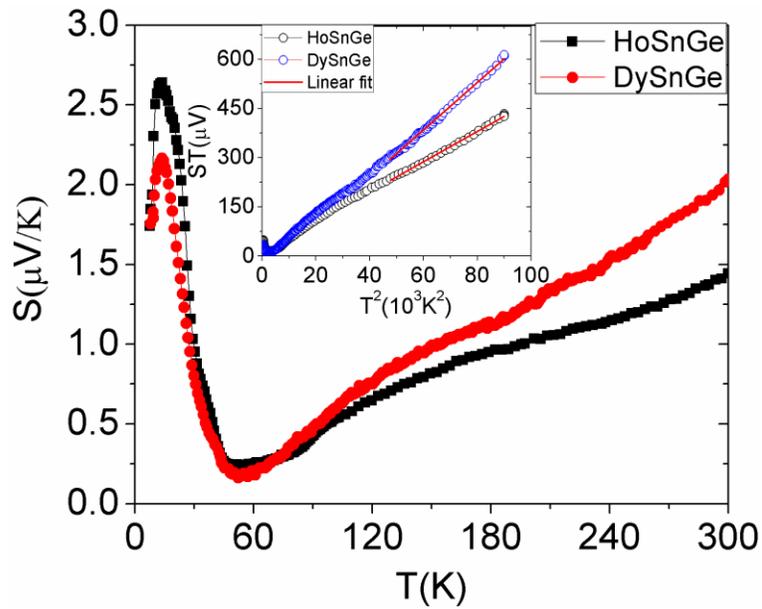

Fig. 9. Temperature variation of the Seebeck coefficient in HoSnGe and DySnGe. The inset shows the fit of the high temperature data.